# Reducing the Complexity of the Linear Programming Decoding

H. Tavakoli, Assistant Professor
Department of Electrical Engineering,
Faculty of Engineering, University of Guilan, Rasht, Iran
htavakoli@guilan.ac.ir

*Abstract*—*In this paper we show how the complexity of Linear Programming (LP) decoder can decrease. We use the degree 3 check equation to model all variation check degrees. The complexity of LP decoding is directed relative to the number of constraint. Number of constraint for original LP decoder is $O(n2^n)$. Our method decrease the number of the constraint to $O(n)$.*

## I. INTRODUCTION

Low density parity check (LDPC) codes were introduced by Gallager [1] in 1963, for the first time, which both iterative decoding structure and constrained random code construction were exploited. Because of increasing the computational ability for computing Gallager's idea, Researchers revisited LDPC codes, and extended the work of Gallager [2-4].

Nowadays, everybody believes that the main subject of recent research in the coding over the digital communication is decoding problem. In decoding problem we interest to find the closest codeword to the received noisy sent codeword. Many decoders such as Sum-Product, Min-Sum and Linear Programming were illustrated for the LDPC codes [5-6].

Linear Programming (LP) decoding is an interested topic for exploring the decoding of LDPC codes. The complexity of the LP decoder is like an original LP problem. So order of complexity for LP decoder is $O(km)$ [7], where k is a constant and m is number of constraint.

In this paper we represent our approach for reducing the complexity of the LP decoder. The rest of this paper organized as follow: in section 2 the original LP decoder will present. In section 3, the main idea for reducing the complexity of LP decoder will present. Conclusion and references present in section 4 and 5.

## II. LP DECODER

LP decoding is a representation of the decoding problem with an optimization problem. In this method, we interest to find the answer of the minimization problem which subjected to relaxed constraint based on the main constraints of the code.

In 2003, LP decoder was presented by Feldman based on his PhD thesis [6]. Feldman illustrated how a decoder can model as a minimization problem. The main formulation for the LP decoder is based on two parts, first the weights of cost function which are based on the received sequence such as $\Gamma = [\gamma_1, \gamma_2, \ldots, \gamma_n]$ and second, the relaxed constraints of a code which are based on the parity check matrix of the code, so we have:

$$\min \Gamma.F \qquad (1)$$

Subject to: $A.F \leq b$.

Our goal in LP decoder is finding the decoded codeword $F = [f_1, f_2, \ldots, f_n]$ which minimize the cost function under the relaxed constraints of the code $A.F \leq b$. The main idea for relaxing the constraints of a code related to Feldman PhD's thesis [6]. Relaxing the constraints for a code is based on choosing an odd set of variables such as S which include $f_i$ from the set of ever parity check constraint such as N and using this formula:

$$\sum_{i \in S} f_i - \sum_{i \in N \setminus S} f_i \leq |S| - 1. \qquad (2)$$

For example for a code with the parity check matrix H as bellow:

$$H = \begin{bmatrix} 1 & 1 & 1 & 0 \\ 0 & 1 & 1 & 1 \end{bmatrix}$$

The problem of the LP decoder is as follows:

$$\min z = \gamma_1 f_1 + \gamma_2 f_2 + \gamma_3 f_3 + \gamma_4 f_4 \qquad (3)$$

Subject to:

$$\begin{bmatrix} +1 & -1 & -1 & 0 \\ -1 & +1 & -1 & 0 \\ -1 & -1 & +1 & 0 \\ +1 & +1 & +1 & 0 \\ 0 & +1 & -1 & -1 \\ 0 & -1 & +1 & -1 \\ 0 & -1 & -1 & +1 \\ 0 & +1 & +1 & +1 \end{bmatrix} \begin{bmatrix} f_1 \\ f_2 \\ f_3 \\ f_4 \end{bmatrix} \leq \begin{bmatrix} 0 \\ 0 \\ 0 \\ +2 \\ 0 \\ 0 \\ 0 \\ +2 \end{bmatrix}, 0 \leq f_i \leq 1 : \forall 1 \leq i \leq 4$$

## III. REDUCING THE COMPLEXITY OF THE LP DECODER

In this part of the paper, we represent a new version of the LP decoder. In this version of the LP decoder, we reduce the number of the relaxed constraints for ever rows of the parity check constraint with ever degree. The main idea of this method is using the constraint for a degree 3 check.

Consider the constraint for a degree 3 check equation $x_1 \oplus x_2 \oplus x_3 = 0$. The relaxed constraint for LP decoder is:

$$+x_1 - x_2 - x_3 \leq 0 \quad (4)$$

$$-x_1 + x_2 - x_3 \leq 0 \quad (5)$$

$$-x_1 - x_2 + x_3 \leq 0 \quad (6)$$

$$+x_1 + x_2 + x_3 \leq 2 \quad (7)$$

This will give us four inequalities. But we have 6 other inequalities which are:

$$0 \leq x_1 \leq 1,\ 0 \leq x_2 \leq 1,\ 0 \leq x_3 \leq 1. \quad (8)$$

But for a check equation of degree 3 has a special property. To see this, add now the (5) and (6) to each other:

$$(-x_1 + x_2 - x_3) + (-x_1 - x_2 + x_3) \leq 0 \Rightarrow -2x_1 \leq 0 \Rightarrow x_1 \geq 0. \quad (9)$$

Now, add the (4) and (7) to each other:

$$(+x_1 - x_2 - x_3) + (+x_1 + x_2 + x_3) \leq 2 \Rightarrow 2x_1 \leq 2 \Rightarrow x_1 \leq 1. \quad (10)$$

This means that the bounding $0 \leq x_1 \leq 1$ of the variables is implied in the equations and 6 inequalities are redundant and we have just 4 inequalities. It is also implied for $x_2, x_3$. If one observes the tetrahedron in the Figure 1 and imagines the four planes, it is possible to see that no solutions are outside the unit cube $[0,1]^3$.

Now, we extend this method for check equation more than degree 3. For a check equation of degree 4, if we use Feldman's relaxation we have:

In general for an LDPC code with n variable and m parity check that the i-th parity check degree is $d_{ci}$ so with this method the $\sum_{j=1}^{m}\left(\sum_{i=0}^{\left[\frac{d_{cj}+1}{2}\right]-1}\binom{d_{cj}}{2i+1}\right) + 2n = \sum_{j=1}^{m} 2^{d_{cj}-1} + 2n$ constraint reduce to $4\sum_{j=1}^{m}(d_{cj}-2)$, we use $\sum_{j=1}^{m}(d_{cj}-3)$ new variable and $\sum_{j=1}^{m}(d_{cj}-2)$ redundant parity check inequalities with degree 3.

With using this method, the number of the relaxed constraints from $O(n2^n)$ decreases to $O(n)$. The Fig.1 illustrates how this method can help us.

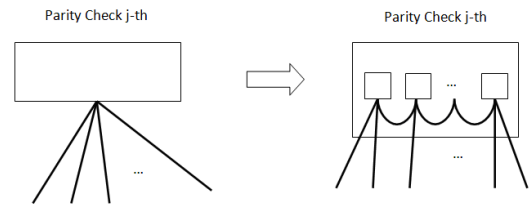

Figure 1. The main idea for reconfiguration of a typical constraint

## IV. CONCLUSION

We propose a method for reducing the complexity of Linear Programming (LP) decoder. This method is effective for implementing in practical processors. Our method presents optimal point of complexity-accuracy for LDPC decoders. We predict other researches, for implementing this method, will represent in the future.